\documentclass{bioinfo}
\copyrightyear{2014}
\pubyear{2014}

\usepackage{hyperref}
\usepackage{url}
\usepackage{amsmath,latexsym,amstext}
\usepackage{mathtools}
\usepackage{algorithm}
\usepackage{algorithmic}
\usepackage{natbib}

\newcommand{\vect}[1]{\boldsymbol{#1}}
\newcommand{\tp}[1]{{#1}^{\mathsf T}}

\newcommand{\eps}{\epsilon}

\renewcommand{\eps}{\varepsilon}
\renewcommand{\epsilon}{\varepsilon}
\renewcommand{\Sigma}{\varSigma}

\DeclareMathAlphabet\mathbfcal{OMS}{cmsy}{b}{n}

\begin{document}
\firstpage{1}

\title[Efficient Bayesian Phylodynamics]{An Efficient Bayesian Inference Framework for Coalescent-Based Nonparametric Phylodynamics}
\author[Shiwei Lan \textit{et~al}]{Shiwei Lan\,$^{1*}$, Julia A. Palacios\,$^{2,3}$, Michael Karcher\,$^{4}$, Vladimir N. Minin\,$^{4,5}$ and Babak Shahbaba\,$^6$\footnote{to whom correspondence should be addressed}}
\address{$^{1}$Department of Statistics, University of Warwick, Coventry CV4 7AL.\\
$^{2}$Department of Ecology and Evolutionary Biology, Brown University.\\
$^{3}$Department of Organismic and Evolutionary Biology, Harvard University.\\
$^{4}$Department of Statistics, University of Washington.\\
$^5$Department of Biology, University of Washington.\\
$^{6}$Department of Statistics, University of California, Irvine.}

\history{}

\editor{}

\maketitle

\begin{abstract}
Phylodynamics focuses on the problem of reconstructing past
population size dynamics from current genetic samples taken from the population of interest. 
This technique has been extensively used in many areas of biology,
but is particularly useful for studying the spread of quickly evolving infectious diseases agents, e.g.,\ influenza virus. 
Phylodynamics inference uses a coalescent model that defines a probability density for 
the genealogy of randomly sampled individuals from the population. When we assume that such a genealogy is known, the coalescent model, equipped with a Gaussian process prior on population size trajectory, allows for nonparametric Bayesian estimation of population size dynamics. While this approach is quite powerful, large data sets collected during infectious disease surveillance challenge the state-of-the-art of Bayesian phylodynamics and demand computationally more efficient inference framework. To satisfy this demand, we provide a computationally efficient Bayesian inference framework based on Hamiltonian Monte Carlo for coalescent process models. Moreover, we show that by splitting the Hamiltonian function we can further improve the efficiency of this approach. Using several simulated and real datasets, we show that our method provides accurate estimates of population size dynamics and is substantially faster than alternative methods based on elliptical slice sampler and Metropolis-adjusted Langevin algorithm.
\end{abstract}

%%%%%%%%%%%%%% Introduction %%%%%%%%%%%%%%%%%%%%

\section{Introduction}
Population genetics theory states that changes in population size affect genetic diversity, leaving a trace of these changes in individuals' genomes. The field of
\emph{phylodynamics} relies on this theory to reconstruct past
population size dynamics from current genetic data. In recent years, phylodynamic inference has become an essential tool in areas like ecology and epidemiology. For example, a study of human influenza A virus from sequences sampled in both hemispheres pointed to a source-sink dynamics of the influenza evolution \citep{Rambaut2008}. 

Phylodynamics connects population dynamics and genetic data using coalescent-based methods \citep{griffiths94,kuhner98,drummond02, strimmer01,drummond05,0pgen-rhein05,heled08,minin08,palacios13}. Typically, phylodynamics relies on Kingman's coalescent model, which is a probability model that describes formation of genealogical relationships of a random sample
of molecular sequences. The coalescent model is parameterized in terms of the  \emph{effective population size}, an indicator of genetic diversity \citep{kingman82}. %Originally developed to estimate stringent parametric forms of a population size trajectory \citep[e.g., exponential growth,][]{griffiths94,kuhner98,drummond02}, recently, coalescent-based methods have been generalized to more flexible nonparametric approaches \citep{strimmer01,drummond05,0pgen-rhein05,heled08,minin08,palacios13}.

While recent studies have shown promising results in alleviating computational difficulties of phylodynamic inference \citep{palacios12, palacios13}, existing methods still lack the level of computational efficiency required to realize the full potential of phylodynamics: \emph{developing surveillance programs that can operate similarly to weather monitoring stations allowing public health workers to predict disease dynamics in order to optimally allocate limited resources in time and space.} To achieve this goal, we present an accurate and computationally efficient inference method for modeling population dynamics given a genealogy. More specifically, we concentrate on a class of Bayesian nonparametric methods based on Gaussian processes
\citep{minin08, gill13, palacios13}. Following \cite{palacios12} and \citet{gill13}, we assume a log-Gaussian process prior on the effective population size. As a result, the estimation of effective population size trajectory becomes similar to the estimation of intensity of a log-Gaussian Cox process \citep[LGCP;][]{moller98}, which is extremely challenging since the likelihood evaluation becomes intractable: it involves integration over an infinite-dimensional random function. We resolve the intractability in likelihood evaluation by discretizing the integration interval with a regular grid to approximate the likelihood and the corresponding score function.

For phylodynamic inference, we propose a computationally efficient Markov chain Monte Carlo (MCMC) algorithm using Hamiltonian Monte Carlo \citep[HMC;][]{duane87,neal10} and one of its variation, called Split HMC \citep{leimkuhler04, neal10,shahbaba13}, which speeds up standard HMC's convergence. Our proposed algorithm has several advantages. First, it updates all model parameters jointly to avoid poor MCMC convergence and slow mixing rates when there are strong dependencies among model parameters \citep{knorr-held02}. Second, unlike a recently proposed Integrated Nested Laplace Approximation \citep[INLA][]{rue09, palacios12} method, our approach can be extended to a more realistic setting 
where instead of a genealogy of the sampled individuals, we observe their genetic data that indirectly inform us about genealogical relationships. Third, we show that our method is up to an order of magnitude more efficient than MCMC algorithms, such as Metropolis-Adjusted Langevin algorithm \citep[MALA;][]{roberts96}, adaptive MALA \citep[aMALA;][]{knorr-held02}, and Elliptical Slice Sampler \citep[ES$^2$;][]{murray10}, 
frequently used in phylodynamics. Finally, although in this paper we focus on phylodynamic studies, our proposed methodology can be easily applied to more general point process models.

The remainder of the paper is organized as follows. In Section 2, we provide a brief overview of the coalescent model and HMC algorithms. Section 3 presents the details of our proposed sampling methods. Experimental results based on simulated and real data are provided in Section 4. Section 5 is devoted to discussion and future directions.

%%%%%%%%%%%%%% Preliminaries %%%%%%%%%%%%%%%%%%%%

\section{Preliminaries}

\subsection{Coalescent}
Assume that a genealogy with time measured in units of generations is available.
The coalescent model allows us to trace the ancestry of a random sample of $n$
genomic sequences as tree: two sequences or lineages merge into a common ancestor as
we go back in time. Those ``merging'' times are called \emph{coalescent times}.
The coalescent with variable population size can be viewed as an inhomogeneous
Markov death process that starts with $n$ lineages at present time, $t_n=0$, and
decreases by one at each of the consequent coalescent times,
$t_{n-1}<\cdots<t_{1}$, until reaching their most recent common ancestor
\citep{griffiths94}.

\begin{figure}[t]
\begin{center}
\includegraphics[width=.48\textwidth]{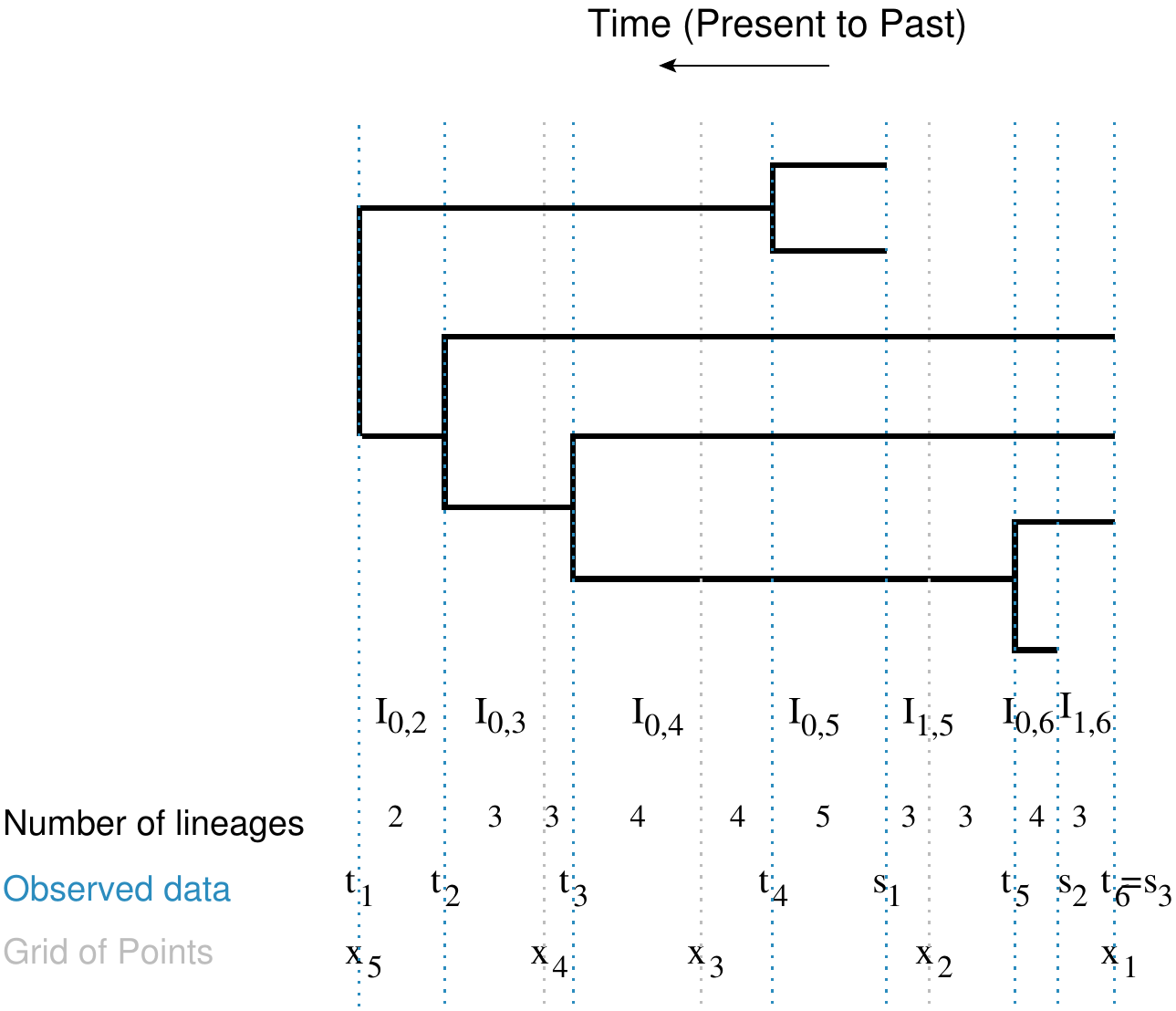}
%I like [scale=.65] better but I am not that obsesed [julia]
\end{center}
\caption{A genealogy with coalescent times and sampling times. Blue dashed lines indicate the observed times: coalescent times $\{t_{1},\cdots,t_{6}\}$ and sampling times $\{s_{1},s_{2},s_{3}\}$. The intervals where the number of lineages change are denoted by $I_{i,k}$. The superimposed grid $\{x_1,\cdots,x_5\}$ is marked by gray dashed lines. We count the number of lineages in each interval defined by grid points, coalescent times and sampling times.}
\label{fig:net}
\end{figure}

Suppose we observe a genealogy of $n$ individuals sampled at time $0$. Under the
standard (\emph{isochronous}) coalescent model, the coalescent times
$t_{n}=0<t_{n-1}<\cdots<t_{1}$, conditioned on the \emph{effective population
size trajectory}, $N_e(t)$, have the density
\begin{equation}\label{isocoal}
\begin{split}
&P[t_{1},\ldots,t_{n} \mid N_{e}(t) ]=\prod^{n}_{k=2}P[t_{k-1} \mid t_{k},N_{e}(t)]\\\
&=\prod^{n}_{k=2}\frac{C_{k}}{N_{e}(t_{k-1})} \exp \left\{ -\int_{I_{k}}\frac{C_{k}}{N_{e}(t)}dt \right\},
\end{split}
\end{equation}
where $C_{k}$ =$ k \choose 2$  and $I_{k}=(t_{k},t_{k-1}]$.
Note that the larger the population size, the longer it takes for two
lineages to coalesce. Further, the larger the number of lineages, the faster two of them
meet their common ancestor \citep{palacios12,palacios13}.

For rapidly evolving organisms, we may have different sampling times. When this
is the case, the standard coalescent model can be generalized to account for such \emph{heterochronous}
sampling \citep{rodrigo99}.
Under the heterochronous coalescent the number of lineages change at both coalescent times and sampling times. Let $\{t_k\}_{k=1}^n$ denote the coalescent times as before, but now let $s_{m}=0<s_{m-1}<\cdots<s_{1}$ denote sampling times of $n_{m},\ldots,n_{1}$ sequences respectively, where $\sum^{m}_{j=1}n_{j}=n$. Further, let $\mathbf{s}$ and $\mathbf{n}$ denote the vectors of sampling times
 $\{s_j\}_{j=1}^m$ and numbers of sequences $\{n_j\}_{j=1}^m$ sampled at these
 times, respectively. Then the coalescent likelihood of a single genealogy becomes
\begin{equation} \label{hetcoal}
\begin{split}
&P[t_{1},\ldots,t_{n} \mid{ \mathbf{s},\mathbf{n}}, N_{e}(t)] = \\
&\prod_{k=2}^{n} \frac{C_{0,k}
\exp \left\{ -\int_{I_{0,k}} \frac{C_{0,k}}{N_{e}(t)}dt -\sum_{i\geq 1}\int_{I_{i,k}} \frac{C_{i,k}}{N_{e}(t)}dt\right\}}
{N_{e}(t_{k-1})},
\end{split}
\end{equation}
where the coalescent factor $C_{i,k}=\binom{l_{i,k}}{2}$ depends on the number of lineages $l_{i,k}$ in the interval $I_{i,k}$ defined by coalescent times and sampling times.
%The factor $C_{0,k}$ depends on the number of lineages $l_{0,k}$ right before a coalescent event occurs at time $t_{k-1}$.
For $k=2,\dots,n$, we denote half-open intervals that end with a coalescent event by
\begin{equation}
I_{0,k}=\left(\left.\max\{t_{k},s_{j}\},t_{k-1}\right.\right],
\end{equation}
for $s_{j}<t_{k-1}$, and half-open intervals that end with a sampling event by ($i>0$)
\begin{equation}
I_{i,k}=\left(\left.\max\{t_{k},s_{j+i}\},s_{j+i-1}\right.\right],
\end{equation}
for $t_{k}<s_{j+i-1}\leq s_j<t_{k-1}$. In density \eqref{hetcoal}, there are $n-1$ intervals $\{I_{i,k}\}_{i=0}$ and $m-1$ intervals $\{I_{i,k}\}_{i>0}$ for all $i,k$. Note that only those intervals satisfying $I_{i,k}\subset (t_{k},t_{k-1}]$ are non-empty. See Figure \ref{fig:net} for more details.

We can think of isochronous coalescence as a special case of heterochronous
coalescence when $m=1, C_{0,k}=C_k, I_{0,k}=I_k, I_{i,k}=\emptyset$ for $i>0$. Therefore, in what follows, we refer to Equation \eqref{hetcoal} as the general case.

We assume the following log-Gaussian Process prior on the effective
population size, $N_{e}(t)$:
\begin{equation}\label{lgp}
N_{e}(t)=\exp[f(t)], \quad f(t)\sim \mathcal{GP}(\mathbf{0},\mathbf{C}(\vect\theta)),
\end{equation}
where $\mathcal{GP}(\mathbf{0},\mathbf{C}(\vect\theta))$ denotes a
Gaussian process with mean function $\mathbf{0}$ and covariance function
$\mathbf{C}(\vect\theta)$. \textit{A priori}, $N_{e}(t)$ is a log-Gaussian Process.

For computational convenience, we use Brownian motion, which is a special case of GP, as our prior for $f(t)$. We define the covariance function as $\mathbf{C}(\kappa) = \frac{1}{\kappa}\mathbf{C}_{BM}$, where the precision parameter $\kappa$ has a $\mathrm{Gamma}(\alpha,\beta)$ prior. For ascending times $0=x_0 <x_1<x_2<\cdots<x_{D-1}$, the $(i,j)$-th element of $\mathbf{C}_{BM}$ is set to $\min\{x_i,x_j\}$. This way, we reduce the computational complexity of inverting the covariance matrix from $\mathcal O(n^3)$ to $\mathcal O(n)$ since the inverse covariance matrix is tri-diagonal \citep{rue05,palacios13}
with elements
\begin{equation*}
\mathbf{C}_{BM}^{-1}(i,j) = \begin{dcases}
\frac{1_{\{i+1\leq D-1\}}}{x_{i+1}-x_i} + \frac{1}{x_i-x_{i-1}}, & \text{ if } i=j,\\
-\frac{1}{|x_i-x_j|}, & \text{ if } |i-j|=1,\\
0, & \text{otherwise}. \end{dcases}
\end{equation*}
In practice we modify the $(1,1)$ element, $1/(x_2-x_1)+1/x_1$, to be $1/(x_2-x_1)$ and denote the resulting precision matrix as ${\bf C}^{-1}_{in}$ to indicate that it comes from an intrinsic autoregression \citep{besag95,knorr-held02}. 
%Matrix ${\bf C}^{-1}_{in}$ is more conservative than ${\bf C}^{-1}_{BM}$ in interval estimate (see experimental results). 
Note that ${\bf C}^{-1}_{in}$ is degenerate in 1 eigen direction so we add a small number, $e$, to the diagonal elements of ${\bf C}^{-1}_{in}$ to make it invertible when ${\bf C}_{in}$ is needed.

\subsection{HMC}
Hamiltonian Monte Carlo \citep{duane87,neal10} is a Metropolis sampling algorithm that suppresses the random walk behavior by proposing states that are distant from the current state, but nevertheless accepts new proposals with high probability. These distant proposals are found by
numerically simulating Hamilton dynamics, whose state space consists of \emph{position}, denoted by the vector $\vect\theta$, and \emph{momentum}, denoted by the vector $\mathbf{p}$. The objective is to sample from the continuous probability distribution with the density function $\pi(\vect\theta)$. It is common to assume $\mathbf{p} \sim \mathcal N(\mathbf{0}, \mathbf{M})$, where $\mathbf{M}$ is a symmetric, positive-definite matrix known as the \emph{mass matrix}, often set to the identity matrix $\mathbf{I}$ for convenience.

In this simulation of Hamiltonian dynamics, the \emph{potential energy}, $U(\vect\theta)$, is defined as the negative log density of $\vect\theta$ (plus any constant); the 
\emph{kinetic energy}, $K(\mathbf{p})$ for momentum variable $\mathbf{p}$, is set to be the negative log density of $\mathbf{p}$ (plus any constant). Then the total energy of the system, the \emph{Hamiltonian} function, is defined as their sum:
$H(\vect\theta, \mathbf{p}) = U(\vect\theta) + K(\mathbf{p})$. Then the system of $(\vect\theta, \mathbf{p})$ evolves according to the following set of \emph{Hamilton's equations}:
\begin{align}
\begin{aligned}
\dot{\vect\theta} & = & \nabla_{\mathbf{p}} H(\vect\theta, \mathbf{p}) & = & \mathbf{M}^{-1}\mathbf{p}, \\
\dot{\mathbf{p}} & = & -\nabla_{\vect\theta} H(\vect\theta, \mathbf{p})& =  & -\nabla_{\vect\theta} U(\vect\theta).
\end{aligned}
\end{align}
In practice, we use a numerical method called \emph{leapfrog} to approximate the Hamilton's equations \citep{neal10} when the analytical solution is not available. We numerically solve the system for $L$ steps, with some step size, $\epsilon$, to propose a new state in the Metropolis algorithm, and accept or reject it according to the Metropolis acceptance probability. \citep[See][for more discussions]{neal10}.

%Split HMC \citep{leimkuhler04,shahbaba13}  improves the computational efficiency of HMC by splitting the Hamiltonian function $H(\vect\theta, \mathbf{p})$ in the following symmetric way:
%\begin{equation}
%H(\vect\theta, \mathbf{p}) = U_1(\vect\theta)/2+[U_0(\vect\theta)+K(\mathbf{p})] + U_1(\vect\theta)/2
%\end{equation}
%where the Hamiltonian dynamics defined by the middle part $H_0(\vect\theta, \mathbf{p}):=U_0(\vect\theta)+K(\mathbf{p})$ is analytically solvable.
%We use the standard leapfrog method for the residual dynamics associated with $U_1(\vect\theta)/2$ for half a step before and after the middle dynamics. 
%If $H_0(\vect\theta, \mathbf{p})$ is a good approximation to $H(\vect\theta, \mathbf{p})$, then simulation of the middle dynamics introduces no error and thus enables the sampler to move faster, saving the computational cost compared to HMC. The Hamiltonian is split in such a symmetric way to ensure the detailed balance condition for the convergence of Markov chain \citep[see more details in][]{shahbaba13}.

%%%%%%%%%%%%%% Method %%%%%%%%%%%%%%%%%%%%

\section{Method}

\subsection{Discretization}\label{sec:discrt}
As discussed above, the likelihood function \eqref{hetcoal} is intractable in general. We can, however, approximate the likelihood using discretization. To this end, we use a fine regular grid, $\mathbf{x}=\{x_d\}_{d=1}^D$, over the observation window and approximate $N_{e}(t)$ by a piece-wise constant function as follows:
\begin{equation}\label{dsct}
N_{e}(t)\approx\sum^{D-1}_{d=1}\exp[f(x^{*}_d)]1_{t \in (x_d,x_{d+1}]}, \quad x^{*}_d=\frac{x_d+x_{d+1}}{2}.
\end{equation}

Note that the regular grid $\mathbf{x}$ does not coincide with the sampling coalescent times, except for the first sampling time $s_m=x_1$ and the last coalescent time $t_1=x_D$.
To rewrite \eqref{hetcoal} using the approximation \eqref{dsct}, we sort all the time points
$\{\mathbf{t},\mathbf{s},\mathbf{x}\}$ to create new $D+m+n-4$
half-open intervals $\{I_{\alpha}^*\}$ with either coalescent time points, sampling
time points, or grid time points as the end points (See Figure \ref{fig:net}).
%We also sort all the time points $\{\mathbf{t},\mathbf{s}\}$ to
%create new $m+n-2$ half-open intervals $I_k$.

For each $\alpha\in\{1,\cdots,D+m+n-4\}$, there exists some $i,k$ and $d$ such that $I_{\alpha}^*= I_{i,k}\cap(x_d,x_{d+1}]$.
Each integral in density \eqref{hetcoal} can be approximated as a sum:
\begin{equation*}
\int_{I_{i,k}} \frac{C_{i,k}}{N_{e}(t)}dt \approx \sum_{I_{\alpha}^*\subset I_{i,k}} C_{i,k}\exp[-f(x^{*}_d)]\Delta_d, 
\end{equation*}
where $\Delta_d:=x_{d+1}-x_d$. This way, the likelihood of coalescent times \eqref{hetcoal} can be rewritten as a product of the following terms:
\begin{equation}
\left\{\frac{C_{i,k}}{\exp[f(x^{*}_d)]}\right\}^{y_{\alpha}} \exp \left\{-\frac{C_{i,k} \Delta_d}{\exp[f(x^{*}_{d})]} \right\},
\end{equation}
where $y_{\alpha}$ is set to 1 if $I_{\alpha}^*$ ends with a coalescent time, and to 0 otherwise.
This happens to be proportional to the probability mass of a Poisson random variable $y_\alpha$ with intensity $\lambda_{\alpha}:=C_{i,k}\Delta_d\exp[-f(x^{*}_{d})]$. Therefore, the likelihood of coalescent times \eqref{hetcoal} can be approximated as follows:
\begin{equation}\label{dsctlik}
\begin{split}
&P[t_{1},\ldots,t_{n} \mid N_{e}(t) ] \approx \prod^{D+m+n-4}_{\alpha=1}P[y_{\alpha} \mid N_{e}(t)]\\
& = \prod_{d=1}^{D-1}\prod_{I_{\alpha}^*\subset(x_d,x_{d+1}]}\left\{\frac{C_{i,k}}{\exp[f(x^{*}_d)]}\right\}^{y_{\alpha}} \exp \left\{-\frac{C_{i,k} \Delta_d}{\exp[f(x^{*}_d)]} \right\},
\end{split}
\end{equation}
where the coalescent factor $C_{i,k}$ on each interval $I_{\alpha}^*$ is determined by the number of lineages $l_{i,k}$ in $I_{\alpha}^*$.

\subsection{Sampling methods}
Denote $\mathbf{f}:=\{f(x_d^*)\}_{d=1}^{D-1}$. Our model can be summarized as
\begin{align}\label{finalModel}
\begin{aligned}
y_{\alpha}|\mathbf{f} & \sim \mathrm{Poisson}[\lambda_{\alpha}(\mathbf{f})],\\
\mathbf{f}|\kappa & \sim \mathcal{N}\left(\mathbf{0}, \frac{1}{\kappa}\mathbf{C}_{in}\right),\\
\kappa & \sim \mathrm{Gamma}(\alpha,\beta).
\end{aligned}
\end{align}
Transforming the coalescent times, sampling times and grid points into $\{y_{\alpha},C_{i,k},\Delta_{d}\}$, we condition on these data to generate posterior samples for
$\mathbf{f}=\log N_e({\bf x}^*)$ and $\kappa$, where ${\bf x}^*=\{x_d^*\}$ is the set of the middle points in \eqref{dsct}. We use these posterior samples to make inference about $N_e(t)$. 

For sampling $\mathbf{f}$ using HMC, we use  \eqref{dsctlik} to compute the discretized
log-likelihood 
\begin{align*}
l &= -\sum_{d=1}^{D-1}\sum_{I_{\alpha}^*\subset(x_d,x_{d+1}]}\left\{y_{\alpha} f(x_d^*)+C_{i,k}\Delta_d\exp[-f(x_d^*)]\right\}
\end{align*}
and the corresponding gradient (score function)
\begin{align*}
s_d & = -\sum_{I_\alpha^*\subset(x_d,x_{d+1}]}\left\{y_{\alpha}-C_{i,k}\Delta_d\exp[-f(x_d^*)]\right\}.
\end{align*}

Because the prior on $\kappa$ is conditionally conjugate, we could directly sample from its full conditional posterior distribution, 
\begin{equation}\label{conj}
\kappa|\cdot \sim \mathrm{Gamma}(\alpha+(D-1)/2,\beta+\mathbf{f}^{\mathsf T}\mathbf{C}_{in}^{-1}\mathbf{f}/2).
\end{equation}
However, updating $\mathbf{f}$ and
$\kappa$ separately is not recommended in general because of their strong interdependency \citep{knorr-held02}: large value of precision $\kappa$ strictly confines the variation of $\mathbf{f}$, rendering slow
movement in the space occupied by $\mathbf{f}$. Therefore, we update $(\mathbf{f},\kappa)$ jointly in MCMC sampling algorithms.
In practice, it is better to sample $\vect\theta:=(\mathbf{f},\tau)$, where $\tau = \log(\kappa)$ is in the same scale as ${\bf f}=\log N_e({\bf x}^*)$. Note that the log-likelihood of $\vect\theta$ is the same as that of $\mathbf{f}$ because equation
 \eqref{hetcoal} does not involve $\tau$. The log-density prior on $\vect\theta$ is defined as follows:
 \begin{equation}
 \log P(\vect\theta) \propto ((D-1)/2+\alpha-1)\tau - (\tp{\mathbf{f}}\mathbf{C}_{in}^{-1}\mathbf{f}/2+\beta)\mathrm e^{\tau}.
 \end{equation}

\subsection{Speed up by splitting Hamiltonian}

Splitting the Hamiltonian is a technique used to speed up HMC
\citep{leimkuhler04,neal10,shahbaba13}. The underlying idea is to divide the total
Hamiltonian into several terms, such that the dynamics associated with some of these terms can be
solved analytically. For these parts, typically quadratic forms, the simulation of the dynamics does not introduce a discretization error, allowing for faster movements in the parameter space. 

We split the Hamiltonian $H(\vect\theta,\mathbf{p})=U(\vect\theta)+K(\mathbf{p})$ as follows:
\begin{equation}\label{splitH}
\begin{split}
& H(\vect\theta, \mathbf{p}) = \frac{-l-[(D-1)/2+\alpha-1]\tau + \beta\mathrm e^{\tau}}{2} +\\
&\frac{\mathbf{f}^{\mathsf T}\mathbf{C}_{in}^{-1}\mathbf{f}\mathrm e^{\tau} +
\tp{\mathbf{p}}\mathbf{p}}{2} + \frac{-l-[(D-1)/2+\alpha-1]\tau + \beta\mathrm e^{\tau}}{2}.
\end{split}
\end{equation}
We further split the middle part (which is the dominant part) into two dynamics involving
$\mathbf{f}|\tau$ and $\tau|\mathbf{f}$ respectively,
\noindent
\begin{subequations}
\hspace{-10pt}\begin{minipage}{.24\textwidth}
\begin{align}\label{HD-f}
\begin{cases}
\begin{aligned}
\dot{\mathbf{f}}|\tau &= \mathbf{p}_{-D},\\
\dot{\mathbf{p}}_{-D} &= -\mathbf{C}_{in}^{-1}\mathbf{f}\mathrm e^{\tau}.
\end{aligned}
\end{cases}
\end{align}
\end{minipage}
\hspace{-7pt}\begin{minipage}{.26\textwidth}
\begin{align}\label{HD-kappa}
\begin{cases}
\begin{aligned}
\dot{\tau}|\mathbf{f} &= p_D,\\
\dot{p}_D &= - \tp{\mathbf{f}}\mathbf{C}_{in}^{-1}\mathbf{f} \mathrm e^{\tau} /2.
\end{aligned}
\end{cases}
\end{align}
\end{minipage}
\end{subequations}
Using the spectral decomposition
$\mathbf{C}_{in}^{-1}=\mathbf{Q}\vect{\Lambda}\mathbf{Q}^{-1}$ and denoting
$\mathbf{f}^*:=\sqrt{\vect{\Lambda}}\mathrm e^{\tau/2}\mathbf{Q}^{-1}\mathbf{f}$ and
$\mathbf{p}^*_{-D}:=\mathbf{Q}^{-1}\mathbf{p}_{-D}$, we can analytically solve the dynamics
\eqref{HD-f} as follows \citep{lan2013} (more details are in the appendix):
\begin{equation*}%\label{rot}
\begin{bmatrix}\mathbf{f}^*(t)\\\mathbf{p}^*_{-D}(t)\end{bmatrix} = 
\begin{bmatrix} \cos(\sqrt{\vect{\Lambda}}\mathrm e^{\tau/2}t) & \sin(\sqrt{\vect{\Lambda}}\mathrm e^{\tau/2}t)\\
-\sin(\sqrt{\vect{\Lambda}}\mathrm e^{\tau/2}t) & \cos(\sqrt{\vect{\Lambda}}\mathrm e^{\tau/2}t)\end{bmatrix}
\begin{bmatrix}\mathbf{f}^*(0)\\\mathbf{p}^*_{-D}(0)\end{bmatrix}.
\end{equation*}
We then use the standard leapfrog method to solve the dynamics \eqref{HD-kappa} and the residual dynamics in
\eqref{splitH} in a symmetric way. Note that we only need to diagonalize $\mathbf{C}_{in}^{-1}$ once before the sampling, and calculate $\tp{\mathbf{f}}\mathbf{C}_{in}^{-1}\mathbf{f} \mathrm e^{\tau} =
\tp{\mathbf{f}^*}\mathbf{f}^*$; therefore, the overall computational complexity of the integrator is $\mathcal O(D^2)$.

%See more details in appendix \ref{apdix:split}.

%Note
%$\tp{\mathbf{f}}\mathbf{C}_{in}^{-1}\mathbf{f} \mathrm e^{\tau} =
%\tp{\mathbf{f}^*}\mathbf{f}^*$. The resulting integrator is as follows:
%\begin{align}
%\mathbf{p}^{(\ell+1/2)} &= \mathbf{p}^{(\ell)} + \eps/2 \begin{bmatrix}\mathbf{s}^{(\ell)}\\ ((D-1)/2+\alpha-1)-\beta\exp(\tau^{(\ell)}) \end{bmatrix}\\
%p_D^{(\ell+1/2)} &= p_D^{(\ell)} - \eps/2\tp{\mathbf{f}^{*(\ell)}}\mathbf{f}^{*(\ell)}/2\\
%\tau^{(\ell+1/2)} &= \tau^{(\ell)} + \eps/2 p_D^{(\ell+1/2)}\\
%\begin{bmatrix}\mathbf{f}^{*(\ell+1)}\\\mathbf{p}^{*(\ell+1/2)}_{-D}\end{bmatrix} &\leftarrow 
%\begin{bmatrix} \cos(\sqrt{\vect{\Lambda}}\exp(\tau^{(\ell+1/2)}/2)\eps) & \sin(\sqrt{\vect{\Lambda}}\exp(\tau^{(\ell+1/2)}/2)\eps)\\
%-\sin(\sqrt{\vect{\Lambda}}\exp(\tau^{(\ell+1/2)}/2)\eps) & \cos(\sqrt{\vect{\Lambda}}\exp(\tau^{(\ell+1/2)}/2)\eps) \end{bmatrix}
%\begin{bmatrix}\mathbf{f}^{*(\ell)}\\\mathbf{p}^{*(\ell+1/2)}_{-D}\end{bmatrix}\\
%\tau^{(\ell+1)} &= \tau^{(\ell+1/2)} + \eps/2 p_D^{(\ell+1/2)}\\
%p_D^{(\ell+1)} &= p_D^{(\ell+1/2)} - \eps/2\tp{\mathbf{f}^{*(\ell+1)}}\mathbf{f}^{*(\ell+1)}/2\\
%\mathbf{p}^{(\ell+1)} &= \mathbf{p}^{(\ell+1/2)} + \eps/2 \begin{bmatrix}\mathbf{s}^{(\ell+1)}\\ ((D-1)/2+\alpha-1) - \beta\exp(\tau^{(\ell+1)}) \end{bmatrix}
%\end{align}
%After applying the integrator $L$ times we get a joint proposal
%$\mathbf{z}^{(L+1)}$ and accept it with probability \eqref{MHacpt}.

\begin{figure}
\begin{center}
\includegraphics[width=.39\textwidth]{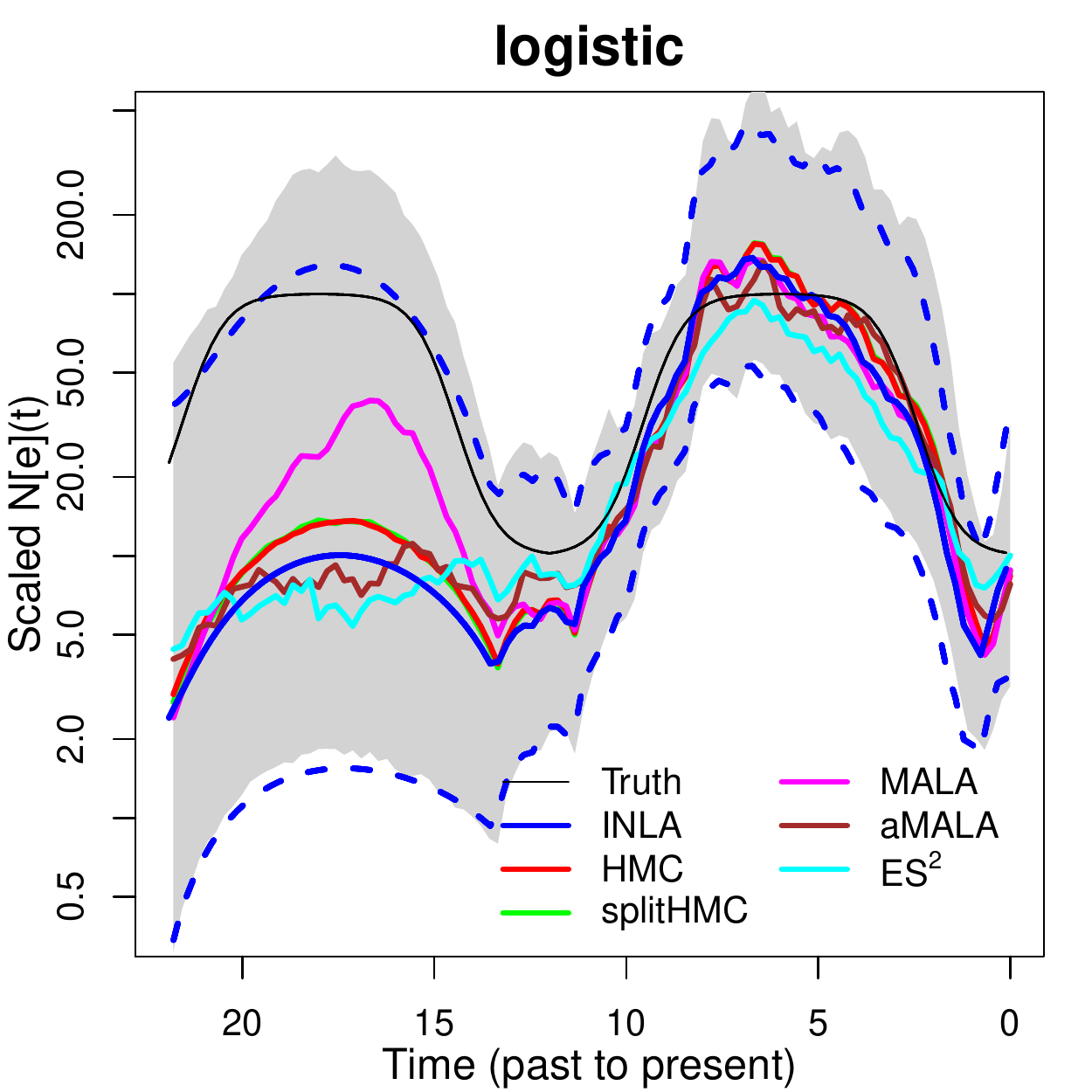}
\includegraphics[width=.39\textwidth]{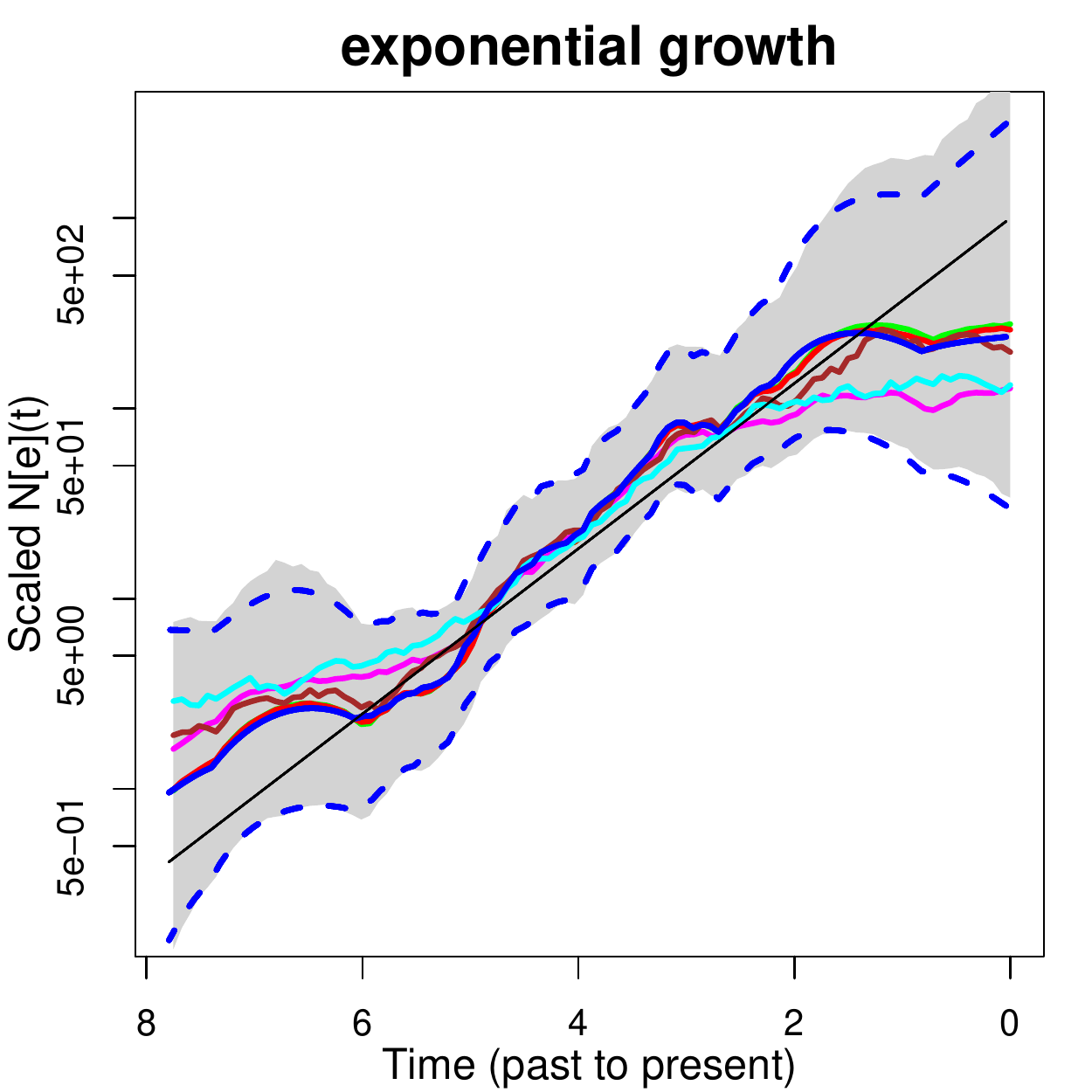}
\includegraphics[width=.39\textwidth]{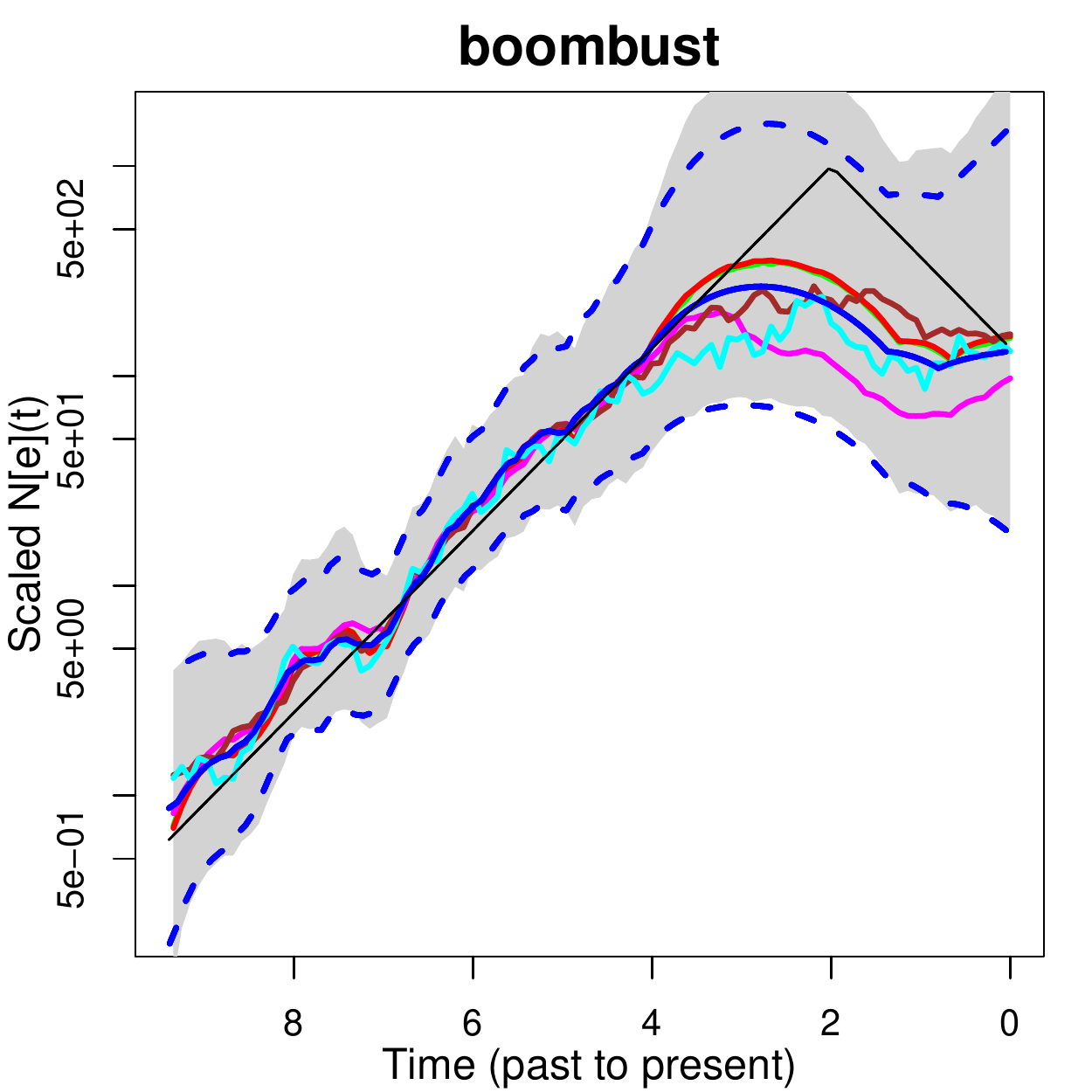}
\end{center}
\vspace{-0.4cm}
\caption{INLA vs MCMC: simulated data under logistic (top), exponential growth (middle) and boombust (bottom) population size trajectories.  Dotted blue lines show $95\%$ credible intervals given by INLA and shaded regions show $95\%$ credible interval estimated with MCMC samples given by splitHMC.}
\label{fig:simulation}
\end{figure}

%%%%%%%%%%%%%% Experiments %%%%%%%%%%%%%%%%%%%%

\section{Experiments}\label{sec:experiments}
We illustrate the advantages of our HMC-based methods using three simulation studies and apply these methods to analysis of a real dataset. We evaluate our methods by comparing them to INLA in terms of accuracy and to several sampling algorithms, MALA, aMALA, and ES$^2$, in terms of sampling efficiency. We define sampling efficiency as time-normalized effective sample size (ESS). Given $B$ MCMC samples for each parameter, we calculate the corresponding ESS =
$B[1 + 2\Sigma_{k=1}^{K}\gamma(k)]^{-1}$, where $\Sigma_{k=1}^{K}\gamma(k)$ is
the sum of $K$ monotone sample autocorrelations \citep{geyer92}. We use the
minimum ESS normalized by the CPU time, s (in seconds), as the overall measure
of efficiency: $\min(\textrm{ESS})/\textrm{s}$.
%In all the experiments discussed here, we set $\alpha=\beta=0.001$ for the precision parameter.

We tune the stepsize and number of leapfrog steps for our HMC-based algorithm such that their overall acceptance probabilities are in a reasonable range (close to 0.70). Since MALA \citep{roberts96} and aMALA \citep{knorr-held02} can be viewed as HMC with one leap frog step for numerically solving Hamiltonian dynamics, we implement MALA and aMALA proposals using our HMC framework. MALA, aMALA, and HMC-based methods update ${\bf f}$ and $\tau$ jointly. aMALA uses a joint block-update method designed for GMRF models: it first generates a proposal 
%$\kappa^*|\kappa\sim p(\kappa^*|\kappa)\propto \frac{\kappa^*+\kappa}{\kappa^*\kappa}$ on $[\kappa/c,\kappa c]$ for some $c>1$.
$\kappa^*|\kappa$ from some symmetric distribution independently of ${\bf f}$, and then updates ${\bf f}^*|{\bf f},\kappa^*$ based on a local Laplace approximation. Finally, $({\bf f}^*,\kappa^*)$ is either accepted or rejected. It can be shown that aMALA is equivalent to a form of Riemannian MALA \citep[][also see Appendix B]{roberts02,girolami11}. In addition, aMALA closely resembles the most frequently used 
MCMC algorithm in Gaussian process-based phylodynamics \citep{minin08,gill13}.

ES$^2$ \citep{murray10} is a commonly used sampling algorithm designed for models with Gaussian process priors. It was also adopted by \citet{palacios13} for phylodynamic inference. ES$^2$ implementation relies on the assumption that the target distribution is approximately normal. This, of course, is not a suitable assumption for the joint distribution of $(\mathbf{f}, \tau)$. Therefore, we alternate the updates $\mathbf{f}|\kappa$ and $\kappa|\mathbf{f}$ when using ES$^2$.

%%%%%%%%%%%%%% Simulations %%%%%%%%%%%%%%%%%%%%

\subsection{Simulations}
We simulate three genealogies relating $n=50$ individuals with the following true trajectories:
\begin{enumerate}
\item logistic trajectory:
\begin{equation*}
N_e(t)=\begin{cases}10+\frac{90}{1+\exp(2(3-(t\!\!\mod 12)))},& t\!\!\mod 12\leq 6,\\
10+\frac{90}{1+\exp(2(-9+(t\!\!\mod 12)))},& t\!\!\mod 12> 6;
\end{cases}
\end{equation*}
\item exponential growth: $N_e(t)=1000\exp(-t)$;
\item boombust:
\begin{equation*}
N_e(t)=\begin{cases}1000\exp(t-2),& t\leq 2,\\
1000\exp(-t+2),& t> 2.
\end{cases}
\end{equation*}
\end{enumerate}
We use $D=100$ equally spaced grid points in the approximation of likelihood when applying INLA and MCMC algorithms (HMC, splitHMC, MALA, aMALA and ES$^2$). 

Figure \ref{fig:simulation} compares the estimates of $N_e(t)$ using INLA and MCMC algorithms in for the three simulations. In general, the results of MCMC algorithms match closely with those of INLA. It is worth noting that MALA and ES$^2$ are occasionally slow to converge. Also, INLA fails when the number of grid points is large, e.g. 10000, while MCMC algorithms can still perform reliably.
% latex table generated in R 3.1.0 by xtable 1.7-3 package
% Thu Jun  5 02:13:42 2014
%\begin{table}[ht]\small
%\centering
%\begin{tabular}{c|l|cccccccc}
%  \hline
%Data & Method & AP & s/iter & minESS($\mathbf{f}$) & minESS($\mathbf{f}$)/s & spdup($\mathbf{f}$) & ESS($\tau$) & ESS($\tau$)/s & spdup($\tau$) \\ 
%  \hline
% & ES$^2$ & 1.00 & 1.53E-03 & 20.17 & 0.88 & 1.00 & 9.97 & 0.43 & 1.00 \\ 
%  I & MALA & 0.81 & 1.87E-03 & 14.88 & 0.53 & 0.60 & 69.95 & 2.49 & 5.72 \\ 
%   & HMC & 0.79 & 8.28E-03 & 303.29 & 2.44 & 2.78 & 418.26 & 3.37 & 7.76 \\ 
%   & splitHMC & 0.70 & 7.14E-03 & 264.92 & 2.47 & 2.82 & 408.09 & 3.81 & 8.77 \\ 
%   \hline
% & ES$^2$ & 1.00 & 1.54E-03 & 19.40 & 0.84 & 1.00 & 12.40 & 0.53 & 1.00 \\ 
%  II & MALA & 0.86 & 1.90E-03 & 16.25 & 0.57 & 0.68 & 54.20 & 1.91 & 3.57 \\ 
%   & HMC & 0.84 & 1.01E-02 & 1113.62 & 7.36 & 8.79 & 585.33 & 3.87 & 7.23 \\ 
%   & splitHMC & 0.74 & 7.66E-03 & 1011.73 & 8.81 & 10.53 & 556.55 & 4.85 & 9.06 \\ 
%   \hline
% & ES$^2$ & 1.00 & 1.53E-03 & 16.30 & 0.71 & 1.00 & 9.98 & 0.44 & 1.00 \\ 
%  III & MALA & 0.91 & 1.90E-03 & 15.02 & 0.52 & 0.74 & 58.55 & 2.05 & 4.69 \\ 
%   & HMC & 0.87 & 1.00E-02 & 660.37 & 4.39 & 6.16 & 558.48 & 3.71 & 8.49 \\ 
%   & splitHMC & 0.76 & 7.66E-03 & 611.25 & 5.32 & 7.48 & 511.95 & 4.46 & 10.20 \\ 
%   \hline
%\end{tabular}
%\caption{Sampling ESSiciency in modeling simulated population trajectories. The true population trajectories are: I) logistic, II) exponential decay, and III) boombust respectively.} 
%\label{simulation}
%\end{table}

%% without MinESS %%
\begin{table}[!t]\scriptsize
\centering
\begin{tabular}{c|l|cccccccc}
\hline
 & Method & AP & s/iter & minESS($\mathbf{f}$)/s & spdup($\mathbf{f}$) & ESS($\tau$)/s & spdup($\tau$) \\ 
  \hline
 & ES$^2$ & 1.00 & 1.56E-03 & 0.22 & 1.00 & 0.25 & 1.00 \\ 
   & MALA & 0.84 & 1.81E-03 & 0.41 & 1.90 & 1.47 & 5.89 \\
 I & aMALA & 0.53 & 4.60E-03 & 0.08 & 0.38 & 0.13 & 0.53 \\  
   & HMC & 0.80 & 9.51E-03 & 1.78 & 8.23 & 1.81 & 7.25 \\ 
   & splitHMC & 0.75 & 7.30E-03 & 2.19 & {\bf 10.13} & 2.51 & {\bf 10.04} \\ 
   \hline
 & ES$^2$ & 1.00 & 1.57E-03 & 0.21 & 1.00 & 0.23 & 1.00 \\ 
   & MALA & 0.78 & 1.82E-03 & 0.32 & 1.53 & 1.14 & 5.03 \\
II & aMALA & 0.53 & 4.61E-03 & 0.09 & 0.41 & 0.18 & 0.79 \\ 
   & HMC & 0.76 & 1.19E-02 & 2.73 & 12.99 & 1.34 & 5.91 \\ 
   & splitHMC & 0.76 & 7.84E-03 & 4.31 & {\bf 20.50} & 2.35 & {\bf 10.40} \\ 
   \hline
 & ES$^2$ & 1.00 & 1.56E-03 & 0.20 & 1.00 & 0.20 & 1.00 \\ 
   & MALA & 0.83 & 1.82E-03 & 0.37 & 1.87 & 1.05 & 5.18 \\
  III & aMALA & 0.53 & 4.61E-03 & 0.08 & 0.40 & 0.14 & 0.67 \\  
   & HMC & 0.81 & 1.18E-02 & 2.22 & 11.24 & 1.17 & 5.75 \\ 
   & splitHMC & 0.72 & 7.41E-03 & 2.87 & {\bf 14.53} & 1.90 & {\bf 9.33} \\ 
   \hline
\end{tabular}
\caption{Sampling efficiency in modeling simulated population trajectories. The true population trajectories are: I) logistic, II) exponential growth, and III) boombust respectively. AP is the acceptance probability. s/iter is the seconds per sampling iteration. ``spdup'' is the speedup of efficiency measurement minESS/s using ES$^2$ as baseline.} 
\label{tab:simulation}
\end{table}

For each experiment we run 15000 iterations with the first 5000 samples discarded. We repeat each experiment 10 times. The results provided in Table \ref{tab:simulation} are averaged over 10 repetitions. As we can see, our methods substantially improve over MALA, aMALA and ES$^2$.
Note that although aMALA has higher ESS compared to MALA, its time-normalized ESS is worse than that of MALA because of its high computational cost of calculating the Fisher information. 

\begin{figure}
\begin{center}
\includegraphics[width=.49\textwidth]{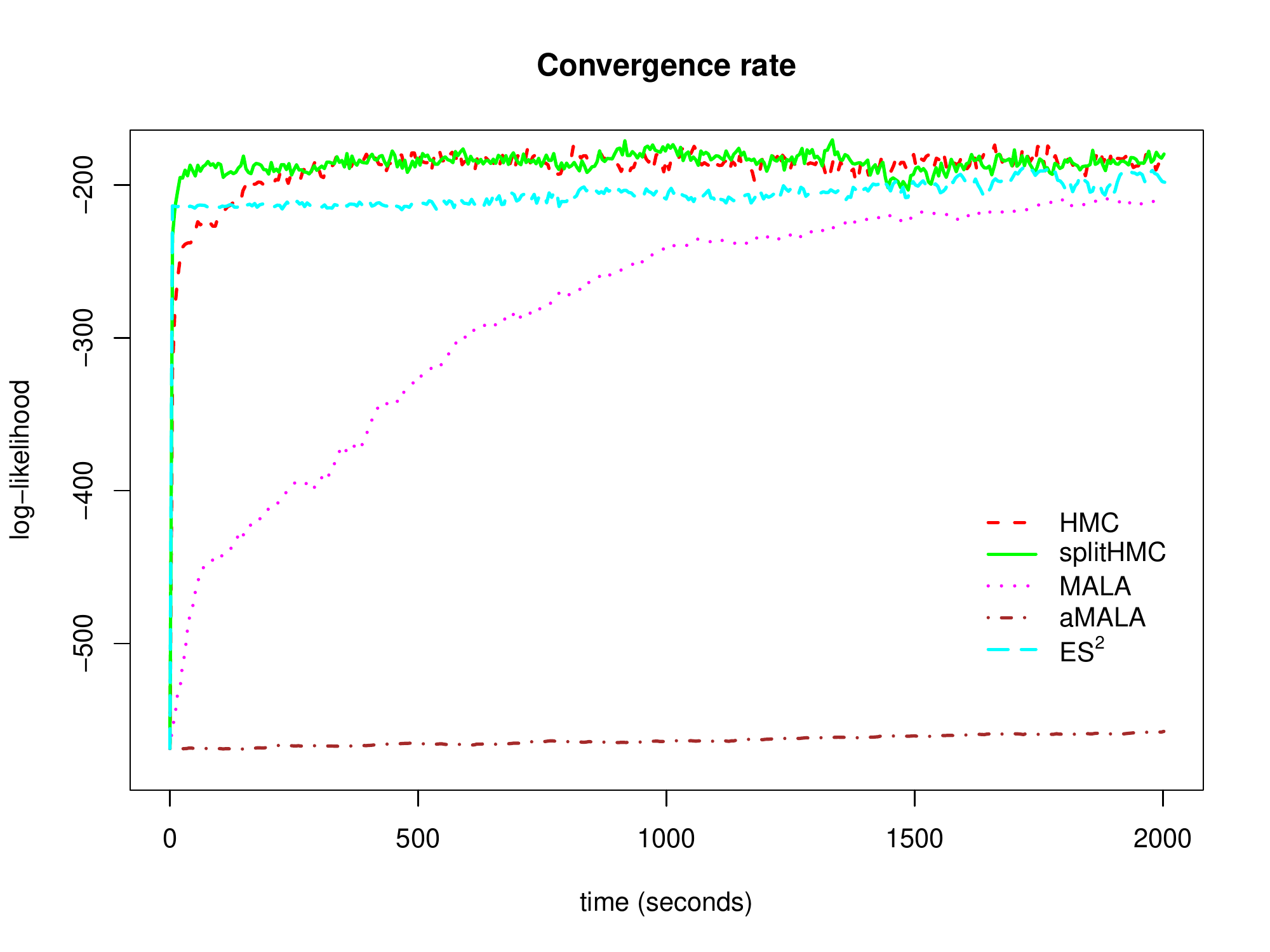}
\end{center}
\caption{Trace plots of log-likelihoods for different sampling algorithms based on a simulated coalescent model with logistic population trajectory. splitHMC converges the fastest.}
\label{fig:mixingrate-fig}
\end{figure}

Figure \ref{fig:mixingrate-fig} compares different sampling methods in terms of their convergence to the stationary distribution when we increase the size of grid points to $D=1000$. As we can see in this more challenging setting, Split HMC has the fastest convergence rate. Neither MALA nor aMALA, on the other hand, has converged within the given time. aMALA is not even getting close to the stationarity, making it much worse than MALA.

%%%%%%%%%%%%%% the first real data %%%%%%%%%%%%%%%%%%%%
\subsection{Human Influenza A in New York}
\begin{figure}
\begin{center}
\includegraphics[width=.45\textwidth]{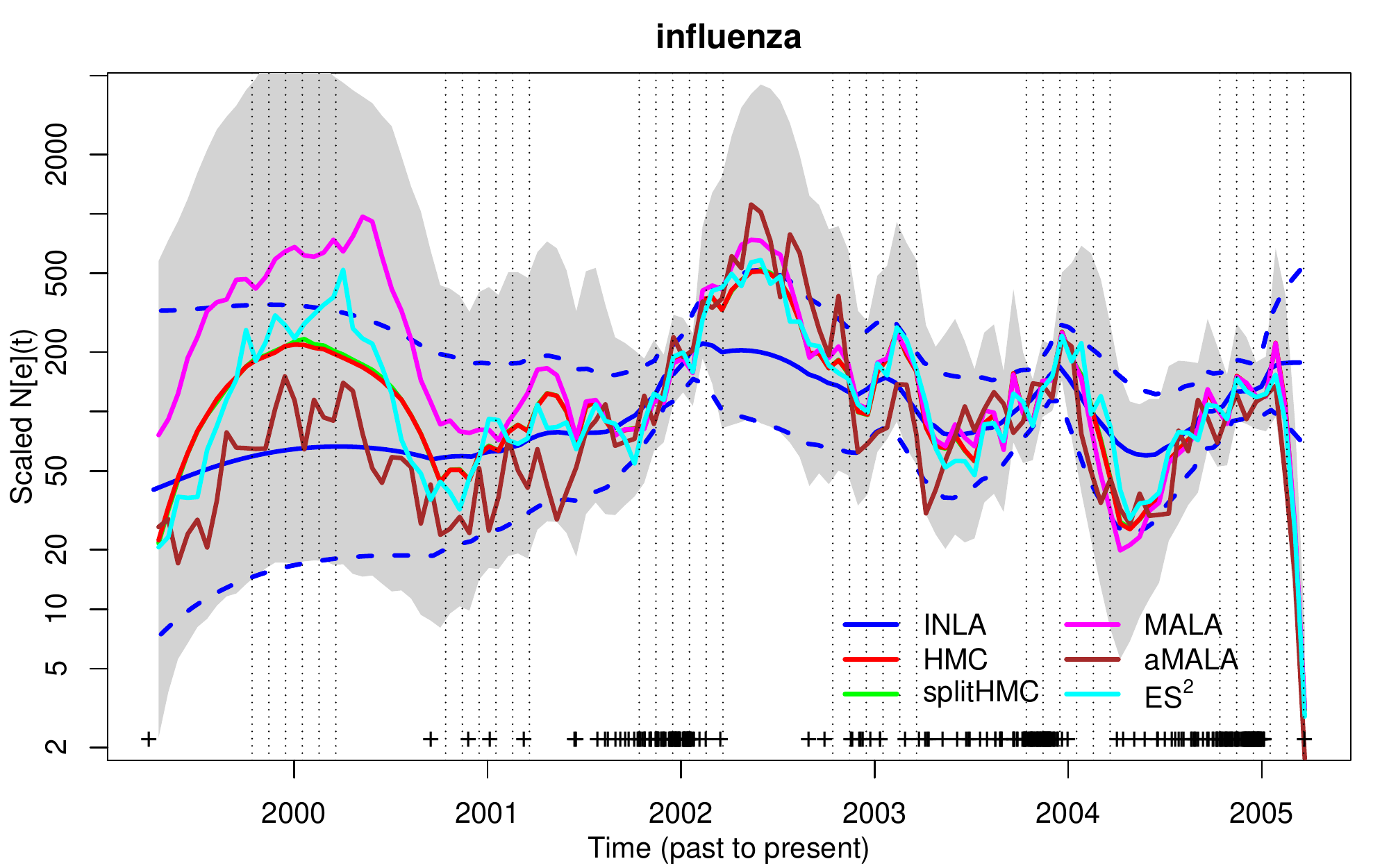}
\includegraphics[width=.45\textwidth]{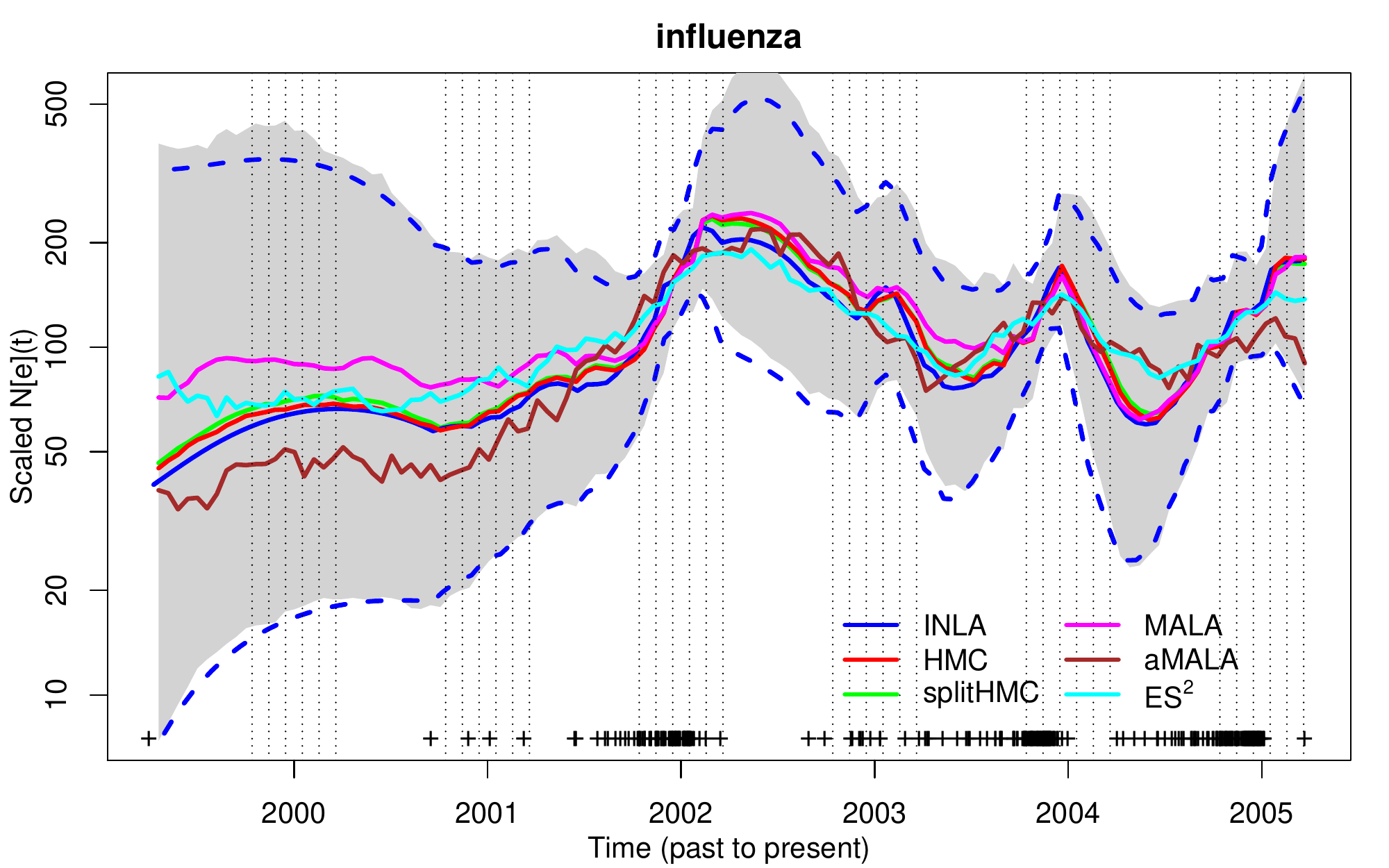}
\end{center}
\caption{Population dynamics of influenza A in New York (2001-2005): shaded region is the $95\%$ credible interval calculated with samples given by splitHMC. Results using ${\bf C}_{BM}^{-1}$ (upper) vs results using ${\bf C}_{in}^{-1}$ (lower).
\label{fig:influenza}}
\end{figure}

Next, we analyze real data based on a genealogy estimated from 288 H3N2 sequences sampled in New York state from January 2001 to March 2005  in order to estimate population size dynamics of
human influenza A in New York \citep{palacios12,palacios13}. The key feature
of the influenza A virus epidemic in temperate regions like New York are the
epidemic peaks during winters followed by strong bottlenecks at the end of the
winter season. We use 120 grid points in the likelihood approximation.
Figure \ref{fig:influenza} shows that with ${\bf C}_{BM}^{-1}$, MCMC algorithms identify such
peak-bottleneck pattern more clearly than INLA. However, their results based on intrinsic precision matrix, ${\bf C}_{in}^{-1}$, are quite comparable to that of INLA. In Table \ref{tab:influenza}, we can see that the speedup by HMC and splitHMC over other MCMC methods is substantial.

% latex table generated in R 3.1.0 by xtable 1.7-3 package
% Wed Jun  4 17:44:53 2014
%\begin{table}[ht]\small
%\centering
%\begin{tabular}{l|cccccccc}
%  \hline
%Method & AP & s/Iter & minESS($\mathbf{f}$) & minESS($\mathbf{f}$)/s & speedup($\mathbf{f}$) & ESS($\tau$) & ESS($\tau$)/s & speedup($\tau$) \\ 
%  \hline
%ES$^2$ & 1.00 & 1.78E-03 & 4.92 & 0.18 & 1.00 & 11.40 & 0.43 & 1.00 \\ 
%  MALA & 0.94 & 2.24E-03 & 11.61 & 0.34 & 1.87 & 52.22 & 1.55 & 3.65 \\ 
%  HMC & 0.91 & 9.62E-03 & 168.81 & 1.17 & 6.37 & 897.54 & 6.22 & 14.61 \\ 
%  splitHMC & 0.83 & 7.94E-03 & 160.85 & 1.35 & 7.35 & 876.92 & 7.36 & 17.30 \\ 
%   \hline
%\end{tabular}
%\caption{Sampling ESSiciency in Influenza data} 
%\label{influenza}
%\end{table}

%% without MinESS %%
\begin{table}[!h]\scriptsize
\centering
\begin{tabular}{l|cccccccc}
\hline
Method & AP & s/Iter & minESS($\mathbf{f}$)/s & spdup($\mathbf{f}$) & ESS($\tau$)/s & spdup($\tau$) \\ 
  \hline
ES$^2$ & 1.00 & 1.92E-03 & 0.34 & 1.00 & 0.27 & 1.00 \\ 
  MALA & 0.78 & 2.12E-03 & 0.29 & 0.86 & 0.60 & 2.20 \\
  aMALA & 0.77 & 8.48E-03 & 0.002 & 0.01 & 0.33 & 1.23 \\ 
  HMC & 0.71 & 1.74E-02 & 1.69 & 4.98 & 0.57 & 2.12 \\ 
  splitHMC & 0.75 & 1.11E-02 & 2.90 & {\bf 8.53} & 1.06 & {\bf 3.92} \\ 
   \hline
\end{tabular}
\caption{Sampling efficiency of MCMC algorithms in influenza data.}
\label{tab:influenza}
\end{table}

%%%%%%%%%%%%%% Discussion %%%%%%%%%%%%%%%%%%%%

\section{Discussion}
In this paper, we have proposed new HMC-based sampling algorithms for phylodynamic inference, which are substantially more efficient than existing methods. Although we have focused on phylodynamics, the methodology presented here can be applied to general LGCP models. 

There are several possible future directions. One possibility is to use ES$^2$ as a proposal generating
mechanism in updating $\mathbf{f}$ as opposed to using it for sampling from the posterior distribution. Finding a good proposal for $\kappa (\tau)$, however, remains challenging. 

An important extension of the methods presented here is to allow for genealogical uncertainty. The MCMC methods analyzed here can be incorporated into a hierarchical framework to infer population size trajectories from sequence data directly. In contrast, INLA cannot be adapted easily to perform inference from sequence data. This greatly limits its generalizability.

%\section*{Acknowledgement}
%Text Text Text Text Text Text  Text Text.  \citealp{Boffelli03} might want to know about  text text text text
%
%\paragraph{Funding\textcolon} Text Text Text Text Text Text  Text Text.

\appendix
\section*{Appendix A: SplitHMC}\label{apdix:split}
Here, we show how to solve Hamiltonian dynamics defined by \eqref{splitH} using the ``splitting'' strategy.
Denote $\mathbf{z}:=(\mathbf{f},\mathbf{p})$, $\mathbf{M}:=\mathbf{C}_{in}^{-1}\mathrm e^{\tau}$, and
$\mathbf{A}:=\begin{bmatrix} \mathbf{0} & \mathbf{I}\\ -\mathbf{M} & \mathbf{0} \end{bmatrix}$.
The dynamics \eqref{HD-f} can be written as
\begin{equation}\label{HDz}
\dot{\mathbf{z}} = \mathbf{A}\mathbf{z}.
\end{equation}
We then have the analytical solution to \eqref{HDz} as
\begin{equation}\label{sol}
\mathbf{z}(t) = \mathrm e^{\mathbf{A}t} \mathbf{z}(0),
\end{equation}
where $\mathrm e^{\mathbf{A}t}$ is a matrix defined as
$\mathrm e^{\mathbf{A}t}:=\sum_{i=0}^{\infty}\frac{t^i}{i!}\mathbf{A}^i$, which in turn can be written as follows:
\begin{equation*}
\begin{split}
\mathrm e^{\mathbf{A}t} &=
\begin{bmatrix}
\mathbf{I} -\frac{t^2}{2!}\mathbf{M}+\frac{t^4}{4!}\mathbf{M}^2+\cdots &  \mathbf{I}t -\frac{t^3}{3!}\mathbf{M} + \frac{t^5}{5!}\mathbf{M}^2+\cdots\\
-\mathbf{M}t +\frac{t^3}{3!}\mathbf{M}^2 -\frac{t^5}{5!}\mathbf{M}^3+\cdots & \mathbf{I} -\frac{t^2}{2!}\mathbf{M}+\frac{t^4}{4!}\mathbf{M}^2+\cdots
\end{bmatrix}\\
&= \begin{bmatrix}
\cos(\sqrt{\mathbf{M}}t) & \mathbf{M}^{-\frac12}\sin(\sqrt{\mathbf{M}}t)\\
-\mathbf{M}^{\frac12}\sin(\sqrt{\mathbf{M}}t) & \cos(\sqrt{\mathbf{M}}t)
\end{bmatrix}\\
&= \begin{bmatrix}
\mathbf{M}^{-\frac12} & \mathbf{0}\\
\mathbf{0} & \mathbf{I}
\end{bmatrix}
\begin{bmatrix}
\cos(\sqrt{\mathbf{M}}t) & \sin(\sqrt{\mathbf{M}}t)\\
-\sin(\sqrt{\mathbf{M}}t) & \cos(\sqrt{\mathbf{M}}t)
\end{bmatrix}
\begin{bmatrix}
\mathbf{M}^{\frac12} & \mathbf{0}\\
\mathbf{0} & \mathbf{I}
\end{bmatrix}.
\end{split}
\end{equation*}
For positive definite matrix $\mathbf{M}$, we can use the spectral decomposition $\mathbf{M}=\mathbf{Q}\mathbf{D}\mathbf{Q}^{-1}$, where $\mathbf{Q}$
is orthogonal matrix, i.e. $\mathbf{Q}^{-1} = \tp{\mathbf{Q}}$. Therefore we have
\begin{equation}
\begin{split}
\mathrm e^{\mathbf{A}t} = &
\begin{bmatrix}
\mathbf{Q} & \mathbf{0}\\
\mathbf{0} & \mathbf{Q}
\end{bmatrix}
\begin{bmatrix}
\mathbf{D}^{-\frac12} & \mathbf{0}\\
\mathbf{0} & \mathbf{I}
\end{bmatrix}
\begin{bmatrix}
\cos(\sqrt{\mathbf{D}}t) & \sin(\sqrt{\mathbf{D}}t)\\
-\sin(\sqrt{\mathbf{D}}t) & \cos(\sqrt{\mathbf{D}}t)
\end{bmatrix}\\
&\cdot\begin{bmatrix}
\mathbf{D}^{\frac12} & \mathbf{0}\\
\mathbf{0} & \mathbf{I}
\end{bmatrix}
\begin{bmatrix}
\mathbf{Q}^{-1} & \mathbf{0}\\
\mathbf{0} & \mathbf{Q}^{-1}
\end{bmatrix}.
\end{split}
\end{equation}
In practice, we only need to diagonalize $\mathbf{C}_{in}^{-1}$ once: $\mathbf{C}_{in}^{-1}=\mathbf{Q}\vect\Lambda\mathbf{Q}^{-1}$, then
$\mathbf{D}=\vect\Lambda\mathrm e^{\tau}$. If we let
$\mathbf{f}^*:=\sqrt{\vect{\Lambda}}\mathrm e^{\tau/2}\mathbf{Q}^{-1}\mathbf{f}$,
$\mathbf{p}^*_{-D}:=\mathbf{Q}^{-1}\mathbf{p}_{-D}$, we have the following solution \eqref{sol}:
\begin{equation*}
\begin{bmatrix}\mathbf{f}^*(t)\\\mathbf{p}^*_{-D}(t)\end{bmatrix} = 
\begin{bmatrix} \cos(\sqrt{\vect{\Lambda}}\mathrm e^{\tau/2}t) & \sin(\sqrt{\vect{\Lambda}}\mathrm e^{\tau/2}t)\\
-\sin(\sqrt{\vect{\Lambda}}\mathrm e^{\tau/2}t) & \cos(\sqrt{\vect{\Lambda}}\mathrm e^{\tau/2}t)\end{bmatrix}
\begin{bmatrix}\mathbf{f}^*(0)\\\mathbf{p}^*_{-D}(0)\end{bmatrix}.
\end{equation*}
We then apply leapfrog method to the remaining dynamics. Algorithm \ref{Alg:splitHMC} summarizes these steps.
\begin{algorithm}[htpb]
\caption{splitHMC for the coalescent model (splitHMC)}
\label{Alg:splitHMC}
\begin{algorithmic}\scriptsize
\STATE Initialize $\vect\theta^{(1)}$ at current $\vect\theta=(\mathbf{f},\tau)$
\STATE Sample a new momentum value $\mathbf{p}^{(1)}\sim \mathcal N(\mathbf{0},\mathbf{I})$
\STATE Calculate $H(\vect\theta^{(1)},\mathbf{p}^{(1)})=U(\vect\theta^{(1)}) + K(\mathbf{p}^{(1)})$ according to \eqref{splitH}
\FOR{$\ell=1$ to $L$}
\STATE
\vspace{-4ex}
\begin{align*}
&\mathbf{p}^{(\ell+1/2)} = \mathbf{p}^{(\ell)} + \eps/2 \begin{bmatrix}\mathbf{s}^{(\ell)}\\ ((D-1)/2+\alpha-1)-\beta\exp(\tau^{(\ell)}) \end{bmatrix}\\
&p_D^{(\ell+1/2)} = p_D^{(\ell)} - \eps/2\tp{\mathbf{f}^{*(\ell)}}\mathbf{f}^{*(\ell)}/2\\
&\tau^{(\ell+1/2)} = \tau^{(\ell)} + \eps/2 p_D^{(\ell+1/2)}\\
&\begin{bmatrix}\mathbf{f}^{*(\ell+1)}\\\mathbf{p}^{*(\ell+1/2)}_{-D}\end{bmatrix} \leftarrow 
\begin{bmatrix} \cos(\sqrt{\vect{\Lambda}}\mathrm e^{\frac12\tau^{(\ell+1/2)}}\eps) & \sin(\sqrt{\vect{\Lambda}}\mathrm e^{\frac12\tau^{(\ell+1/2)}}\eps)\\
-\sin(\sqrt{\vect{\Lambda}}\mathrm e^{\frac12\tau^{(\ell+1/2)}}\eps) & \cos(\sqrt{\vect{\Lambda}}\mathrm e^{\frac12\tau^{(\ell+1/2)}}\eps) \end{bmatrix}\\
&\phantom{\leftarrow}\cdot\begin{bmatrix}\mathbf{f}^{*(\ell)}\\\mathbf{p}^{*(\ell+1/2)}_{-D}\end{bmatrix}\\
&\tau^{(\ell+1)} = \tau^{(\ell+1/2)} + \eps/2 p_D^{(\ell+1/2)}\\
&p_D^{(\ell+1)} = p_D^{(\ell+1/2)} - \eps/2\tp{\mathbf{f}^{*(\ell+1)}}\mathbf{f}^{*(\ell+1)}/2\\
&\mathbf{p}^{(\ell+1)} = \mathbf{p}^{(\ell+1/2)} + \eps/2 \begin{bmatrix}\mathbf{s}^{(\ell+1)}\\ ((D-1)/2+\alpha-1) - \beta\exp(\tau^{(\ell+1)}) \end{bmatrix}
\end{align*}
\ENDFOR
\STATE Calculate $H(\vect\theta^{(+1)},\mathbf{p}^{(L+1)})=U(\vect\theta^{(L+1)}) + K(\mathbf{p}^{(L+1)})$ according to \eqref{splitH}
\STATE Calculate the acceptance probability $\alpha =\min\{1, \exp[-H(\vect\theta^{(+1)},\mathbf{p}^{(L+1)})+H(\vect\theta^{(1)},\mathbf{p}^{(1)})] \}$
\STATE Accept or reject the proposal according to $\alpha$ for the next state $\vect\theta'$
\end{algorithmic}
\end{algorithm}
\begin{algorithm}[htpb]
\caption{Adaptive MALA (aMALA)}
\label{Alg:aMALA}
\begin{algorithmic}
\STATE Given current state $\vect\theta=(\mathbf{f},\kappa)$ calculate potential energy $U(\vect\theta)$
\REPEAT
\STATE $z\sim \mathrm{Unif}[1/c,c]$,\; $u\sim \mathrm{Unif}[0,1]$
\UNTIL{$u<\frac{z+1/z}{c+1/c}$}
\STATE update precision parameter $\kappa^*=\kappa z$
\STATE Sample momentum ${\bf p}\sim \mathcal N({\bf 0},{\bf G}({\bf f},\kappa^*)^{-1})$
\STATE Calculate log of proposal density $\log p({\bf f}^*|{\bf f},\kappa^*)=-\frac{1}{2}\tp{\bf p}{\bf G}({\bf f},\kappa^*){\bf p} + \frac{1}{2}\log\det{\bf G}({\bf f},\kappa^*)$
\STATE update momentum ${\bf p} \leftarrow {\bf p} -\eps/2 {\bf G}({\bf f},\kappa^*)^{-1}\nabla U({\bf f},\kappa^*)$
\STATE update latent variables ${\bf f}^*={\bf f} + \eps {\bf p}$
\STATE update momentum ${\bf p} \leftarrow {\bf p} -\eps/2 {\bf G}({\bf f}^*,\kappa)^{-1}\nabla U({\bf f}^*,\kappa)$
\STATE Calculate log of reverse proposal density $\log p({\bf f}|{\bf f}^*,\kappa)=-\frac{1}{2}\tp{\bf p}{\bf G}({\bf f}^*,\kappa){\bf p} + \frac{1}{2}\log\det{\bf G}({\bf f}^*,\kappa)$ \STATE Calculate new potential energy $U(\vect\theta^*)$
\STATE Accept/reject the proposal according to $\log\alpha=-U(\vect\theta^*)+U(\vect\theta)-\log p({\bf f}^*|{\bf f},\kappa)+\log p({\bf f}|{\bf f}^*,\kappa)$ for the next state $\vect\theta'$
\end{algorithmic}
\end{algorithm}

\section*{Appendix B: Adaptive MALA}\label{apdix:aMALA}
We now show that the joint block updating in \cite{knorr-held02} can be recognized as an adaptive MALA algorithm. First, we sample $\kappa^*|\kappa\sim p(\kappa^*|\kappa)\propto \frac{\kappa^*+\kappa}{\kappa^*\kappa}$ on $[\kappa/c,\kappa c]$ for some $c>1$ controlling the step size of $\kappa$.
Denote ${\bf w}:=\{ C_{i,k} \Delta_d\}_{1}^{D+m+n-4}$ and use the following Taylor expansion for $\log p({\bf f}|\kappa)$ about $\hat{\bf f}$:
\begin{equation*}
\begin{split}
&\log p({\bf f}|\kappa) = -\tp{\bf y}{\bf f} -\tp{\bf w}\exp(-{\bf f}) -\frac{1}{2}\tp{\bf f} \kappa\mathbf{C}_{in}^{-1}{\bf f}\\
&\approx -\tp{\bf y}{\bf f} -\tp{({\bf w}\exp(-\hat{\bf f}))}[{\bf 1}-({\bf f}-\hat{\bf f}) + ({\bf f}-\hat{\bf f})^2/2] -\frac{1}{2}\tp{\bf f} \kappa\mathbf{C}_{in}^{-1}{\bf f}\\
&= \tp{(-{\bf y}+{\bf w}\exp(-\hat{\bf f})({\bf 1}+\hat{\bf f}))}{\bf f} + \frac{1}{2}\tp{\bf f}[\kappa\mathbf{C}_{in}^{-1} + \mathrm{diag}({\bf w}\exp(-\hat{\bf f}))] {\bf f}\\
&=:\tp{\bf b}{\bf f} - \frac{1}{2} \tp{\bf f}{\bf G}{\bf f},
\end{split}
\end{equation*}
where ${\bf b}(\hat{\bf f}):=-{\bf y}+{\bf w}\exp(-\hat{\bf f})({\bf 1}+\hat{\bf f})$, ${\bf G}(\hat{\bf f},\kappa):=\kappa\mathbf{C}_{in}^{-1} + \mathrm{diag}({\bf w}\exp(-\hat{\bf f}))$.
Setting $\hat{\bf f}$ to the current state, ${\bf f}$, and propose ${\bf f}^*|{\bf f},\kappa^*$ from the following Gaussian distribution:
\begin{equation*}
{\bf f}^*|{\bf f},\kappa^* \sim \mathcal N(\vect\mu, \vect\Sigma),
\end{equation*}
with
\[
\begin{split}
\vect\mu &= {\bf G}({\bf f},\kappa^*)^{-1}{\bf b}({\bf f}) = {\bf f} + {\bf G}({\bf f},\kappa^*)^{-1} \nabla_{\bf f} \log p({\bf f}|\kappa^*) \text{ and }\\
\vect\Sigma &= {\bf G}({\bf f},\kappa^*)^{-1},
\end{split}
\]
which has the same form as Langevin dynamical proposals. Interestingly, ${\bf G}(\hat{\bf f},\kappa)$ is exactly the (observed) Fisher information. That is, this approach is equivalent to Riemannian MALA \citep{girolami11}.

Finally, $\vect\theta^*=({\bf f}^*,\kappa^*)$ is jointly accepted with the following probability:
\begin{equation*}
\begin{split}
\alpha &= \min\left\{1,\frac{p(\vect\theta^*|\mathcal D)}{p(\vect\theta|\mathcal D)} \frac{p(\kappa|\kappa^*)p({\bf f}|{\bf f}^*,\kappa)}{p(\kappa^*|\kappa)p({\bf f}^*|{\bf f},\kappa^*)}\right\}\\
& = \min\left\{1,\frac{p(\vect\theta^*|\mathcal D)}{p(\vect\theta|\mathcal D)} \frac{p({\bf f}|{\bf f}^*,\kappa)}{p({\bf f}^*|{\bf f},\kappa^*)}\right\},
\end{split}
\end{equation*}
where $p(\kappa^*|\kappa)$ is a symmetric proposal. Algorithm \ref{Alg:aMALA} summarizes the steps for adaptive MALA.

\newpage
\twocolumn[
\bibliographystyle{plainnat}
\bibliography{references}
]

\end{document}